# A suggested interpretation of the quantum theory in terms of discontinuous motion of particles

Rui Qi

Institute of Electronics, Chinese Academy of Sciences

17 Zhongguancun Rd., Beijing, China

E-mail: rg@mail.ie.ac.cn

We present a theory of discontinuous motion of particles in continuous space-time. We show that the simplest nonrelativistic evolution equation of such motion is just the Schrödinger equation in quantum mechanics. This strongly implies what quantum mechanics describes is discontinuous motion of particles. Considering the fact that space-time may be essentially discrete when considering gravity, we further present a theory of discontinuous motion of particles in discrete space-time. We show that its evolution will naturally result in the dynamical collapse process of the wave function, and this collapse will bring about the appearance of continuous motion of objects in the macroscopic world.

# Introduction

As we know, what classical mechanics describes is continuous motion of particles. Then a natural question appears when we turn to quantum mechanics, i.e. which motion of particles does quantum mechanics describe[1]? But unfortunately this is not an easy question. In fact, it is a hard problem, and people have been arguing with each other about its solution since the founding of quantum mechanics[1-6]. In this paper, we will try to solve this problem along a clear logical way. A convincing solution will be found in the end of the road.

The plan of this paper is as follows: In Sect. 2 we first denote that what quantum mechanics describes is not continuous motion of particles. As an example, we re-analyze the famous double-slit experiment. In Sect. 3 we present a theory of discontinuous motion of particles in continuous space-time. We show that the simplest evolution law of discontinuous motion is just the Schrödinger equation in quantum mechanics. This further implies what quantum mechanics describes is discontinuous motion of particles. In Sect. 4 we try to interpret the theory of discontinuous motion. Two alternatives are given, and we demonstrate that the existence of collapse of wave function may be inevitable. In Sect. 5 we point out that space-time may be essentially discrete when considering the proper combination of quantum mechanics and general relativity, and give a simple demonstration. In Sect. 6 we discuss the influences of discreteness of space-time to the evolution of discontinuous motion. The possible evolution law of the discontinuous motion of particles in discrete space-time is worked out, and we demonstrate that it will naturally result in the dynamical collapse process of the wave function. In Sect. 7 we further show that continuous motion and its evolution law can be consistently derived from the evolution law of the discontinuous motion in discrete space-time. Conclusions are given in Sect. 8.

---

[1] In this paper we assume that there exists only one kind of physical reality---particles.

# What quantum mechanics describes is not continuous motion of particles

Even though people haven't known what does quantum mechanics describe yet, they indeed know what does quantum mechanics not describe. It is well known that quantum mechanics doesn't describe continuous motion of particles, or we can say, what quantum mechanics describes is not continuous motion of particles. Here as an example, let's have a look at the well-known double-slit experiment, and see why quantum mechanics doesn't describe continuous motion of particles.

In the usual double-slit experiment, the single particle such as photon is emitted from the source one after the other, and then passes through the two slits to arrive at the screen. In this way, when a large number of particles reach the screen, they form the double-slit interference pattern. Now we will demonstrate that this experiment clearly reveals what quantum mechanics describes is not continuous motion of particles. Using apagoge, if the motion of particle is continuous, then the particle can only pass through one of the two slits, and it is not influenced by the other slit in each experiment. Thus it is evident that the double-slit interference pattern will be the same as the direct mixture of two one-slit patterns, each of which is formed by opening each of the two slits, since the passing process of each particle in double-slit experiment is exactly the same as that in one of the two one-slit experiments. But quantum mechanics predicts that there exist obvious differences between the interference patterns of the above two situations, and all known experiments coincide with the prediction. Thus the motion of particle described by quantum mechanics can't be continuous, and the particle must pass through both slits in some unusual way during passing through the two slits.

Now there appears a simple but subtle question, i.e. if the motion of the particles described by quantum mechanics is not continuous, then which form is it? If there exists only one kind of physical reality---particles, and the objective motion picture of the particles can't be essentially rejected, then the motion of particles must be discontinuous. This is an inevitable logical conclusion. As we think, this answer is more direct and natural, since classical mechanics describes continuous motion, then correspondingly quantum mechanics will describe another different kind of motion, namely discontinuous motion. But the answer seems very bizarre, and we have never learned the discontinuous motion. Now let's be close to it and grasp it.

## A theory of discontinuous motion of particles

In this section, we will present a theory of discontinuous motion of particles. Our analyses will show that the simplest evolution law of discontinuous motion of particles is just the Schrödinger equation in quantum mechanics. This strongly implies what quantum mechanics describes is discontinuous motion of particles.

## A general analysis

First, we will strictly define the discontinuous motion of particles using three presuppositions about the relation between physical motion and mathematical point set. They are the basic conceptions and correspondence rules needed before we discuss the discontinuous motion of particles in continuous space-time.

(1). Time and space in which the particle moves are both continuous.

(2). The moving particle is represented by one point in time and space.

(3). The discontinuous motion of particle is represented by the dense point set in time and space.

The first presupposition defines the continuity of space-time. The second one defines the existent form of particle in time and space. The last one defines the discontinuous motion of particle using the mathematical point set. For simplicity but without losing generality, in the following we will mainly analyze the point set in two-dimensional space-time, which corresponds to one-dimensional motion in continuous space-time.

We first introduce what is dense point set. As we know, the point set theory has been deeply studied since the beginning of the 20th century. Nowadays we can grasp it more easily. According to this theory, we know that the general point set is dense point set, whose basic property is the measure of the point set. Its visualizing picture is like a mass of fog or cloud. While the continuous point set is the familiar curve, one kind of special dense point set, and its basic property is the length of the point set. It is indeed a wonder that so many points bind together to form a continuous curve by order.

Enlightened by the theory of fluid mechanics we can find the description of the dense point set, which corresponds to the discontinuous motion of particles in continuous space-time. The mathematical analysis shows that the proper description of the dense point set, or the motion state of a particle undergoing the discontinuous motion is position measure density $\rho(x,t)$ and position measure flux density $j(x,t)$, and they satisfy the measure conservation equation $\frac{\partial \rho(x,t)}{\partial t} + \frac{\partial j(x,t)}{\partial x} = 0$. From the above correspondence rules, we can clearly see the physical meaning of the description quantities $\rho(x,t)$ and $j(x,t)$. As to the position measure density $\rho(x,t)$, it represents the relative frequency of the particle appearing in the infinitesimal space interval $dx$ near the position $x$ during the infinitesimal interval $dt$ near the instant $t$, and we can measure it through directly measuring the appearing probability of the particle in the above situation. Thus $\rho(x,t)$ possesses a direct physical meaning. However, the position measure flux density $j(x,t)$ possesses no direct physical meaning, and we can only measure it through indirect measurement.

It is very natural to extend the basic descriptions of the motion of a single particle to the many particles situation. As to the motion state of N particles, we can define their joint position measure density $\rho(x_1, x_2, ... x_N, t)$ according to the theory of point set, it represents the appearing probability of the situation, in which particle 1 is in position $x_1$, particle 2 is in position $x_2$, ...and particle N is in position $x_N$. In a similar way, we can define the joint position measure flux density $j(x_1, x_2,...x_N, t)$. It satisfies the joint measure conservation equation: $\frac{\partial \mathbf{r}(x_1, x_2,...x_N, t)}{\partial t} + \sum_{i=1}^{N} \frac{\partial j(x_1, x_2,...x_N, t)}{\partial x_i} = 0$. We can easily see that, the descriptions of the motion of many particles, namely the joint position measure density $\mathbf{r}(x_1, x_2,...x_N, t)$ and joint position measure flux density $j(x_1, x_2,...x_N, t)$ are naturally defined in the 3N dimensional configure space, not in the real space. Besides, when the N particles are independent, the joint position measure density $\mathbf{r}(x_1, x_2,...x_N, t)$ can be reduced to the product of the position measure density of each particle, namely $\mathbf{r}(x_1, x_2,...x_N, t) = \prod_{i=1}^{N} \mathbf{r}(x_i, t)$.

### The evolution law

In the following, we will try to find the possible evolution equations of the discontinuous motion of particles. Here we mainly analyze one-dimensional motion, but the results can be easily extended to the three-dimensional situation.

First, we need to find the simplest solution of the evolution equation, in which we can find the first motion principle similar to Newton's first principle. It is evident that the simplest solution of the motion equation is:

$$\frac{\partial \mathbf{r}(x,t)}{\partial t} = 0 \quad \text{------ (1)}$$

$$\frac{\partial j(x,t)}{\partial t} = 0 \quad \text{------ (2)}$$

$$\frac{\partial \mathbf{r}(x,t)}{\partial x} = 0 \quad \text{------ (3)}$$

$$\frac{\partial j(x,t)}{\partial x} = 0 \quad \text{------ (4)}$$

using the relation $j = \mathbf{r}v$ we can further get the solution: $\mathbf{r}(x,t) = 1$, $j(x,t) = v = p/m$, where $m$ is the mass of the particle, and $p$ is defined as the momentum of the particle.

Now we get the first motion principle, i.e. during the free motion of particle, the momentum of the particle is invariant. It can be easily seen that, contrary to continuous motion, for the free particle with one constant momentum, its position will not be limited in the infinitesimal space interval $dx$, but spread throughout the whole space with the same position measure density.

Similar to the quantity position, the natural assumption in logic is also that the momentum (motion) state of a particle in infinitesimal interval $dt$ is still a general dense point set in momentum space. Thus we can also define the momentum measure density $f(p,t)$, which satisfies the normalization relation $\int_{-\infty}^{+\infty} f(p,t)dp = 1$[1], and the momentum measure flux density $J(p,t)$. Their meanings are similar to those of position, and satisfy the similar measure conservation equation
$$\frac{\partial f(p,t)}{\partial t} + \frac{\partial J(p,t)}{\partial p} = 0.$$

Then we have two kinds of description quantities---one is position, the other is momentum. Position descriptions $\mathbf{r}(x,t)$ and $j(x,t)$ provide a complete local description of the motion state. This we may call the local description of the discontinuous motion. Similarly momentum descriptions $f(p,t)$ and $J(p,t)$ provide a complete nonlocal description of the motion state. For a particle with any constant momentum, its position will spread throughout the whole space with the same position measure density. This we may call the nonlocal description of the discontinuous motion. Since at any instant the motion state of a particle is unique, there should exist a one-to-one relation between these two kinds of descriptions, i.e. there should exist a one-to-one relation between position description $r(x,t)$, $j(x,t)$ and momentum description $f(p,t)$, $J(p,t)$, and this relation is irrelevant to the concrete motion state. In the following we will find the one-to-one relation, and our analysis will also show that this relation essentially determines the possible evolution of motion.

---

[1] For some ideal situations where the integrals $\int_{-\infty}^{+\infty} \mathbf{r}(x,t)dx$ and $\int_{-\infty}^{+\infty} f(p,t)dp$ turn to be infinite, the general normalization relation will be $\int_{-\infty}^{+\infty} \mathbf{r}(x,t)dx = \int_{-\infty}^{+\infty} f(p,t)dp$.

It is evident that there exists no direct one-to-one relation between the measure density functions $r(x,t)$ and $f(p,t)$, since even for the constant momentum situation, we have $r(x,t)=1$ and $f(p,t)=d^2(p-p_0)$, and there is no one-to-one relation between them. Then in order to obtain the one-to-one relation, we have to construct a new kind of integrative description on the basis of the above position description $r(x,t)$, $j(x,t)$ and momentum description $f(p,t)$, $J(p,t)$. Here we only give the main clues and the detailed mathematical demonstrations are omitted.

First, we disregard the time variable $t$ or let $t=0$. As to the above free evolution state with one momentum, we have $r(x,0)=1$, $j(x,0)=p_0/m$ and $f(p,0)=d^2(p-p_0)$, $J(p,0)=0$. Thus we need to synthesize a new position state function $y(x,0)$ using 1 and $p_0/m$, and a new momentum state function $j(p,0)$ using $\boldsymbol{d}^2(p-p_0)$ and 0, and find the one-to-one relation between these two state functions. We generally write it as follows:

$$\boldsymbol{y}(x,0) = \int_{-\infty}^{+\infty} \boldsymbol{j}(p,0) T(p,x) dp \quad \text{------ (5)}$$

where $T(p,x)$ is the transformation function and is generally continuous and finite for finite $p$ and $x$.

Since the function $\boldsymbol{j}(p,0)$ will contain some form of the basic element $\boldsymbol{d}^2(p-p_0)$, normally we may expand it as $\boldsymbol{j}(p,0) = \sum_{i=1}^{\infty} a_i \boldsymbol{d}^i(p-p_0)$. Besides, the function $\boldsymbol{y}(x,0)$ will contain the momentum $p_0$, and be generally continuous and finite for finite $x$. Then it is evident that the function $\boldsymbol{j}(p,0)$ can only contain the term $\boldsymbol{d}(p-p_0)$, because the other terms will result in infiniteness. Furthermore, the result $\boldsymbol{j}(p,0) = \boldsymbol{d}(p-p_0)$ implies that there exists the simplest relation $f(p,0) = \boldsymbol{j}^*(p,0)\boldsymbol{j}(p,0)$ [1], and owing to the equality between the position description and momentum description, we also have the similar relation $\boldsymbol{r}(x,0) = \boldsymbol{y}^*(x,0)\boldsymbol{y}(x,0)$.

Then we may let $\boldsymbol{y}(x,0) = e^{iG(p_0,x)}$ and have $T(p,x) = e^{iG(p,x)}$. Considering the symmetry between the properties position and momentum [2], we have the general extension $G(p,x) = \sum_{i=1}^{\infty} b_i (px)^i$. Furthermore, this kind of symmetry also results in the symmetry between the transformation $T(p,x)$ and its reverse transformation $T^{-1}(p,x)$, where $T^{-1}(p,x)$ satisfies the relation:

$$\boldsymbol{j}(p,0) = \int_{-\infty}^{+\infty} \boldsymbol{y}(x,0) T^{-1}(p,x) dx \quad \text{------ (6)}$$

---

[1] Evidently, another simple relation $f(p,0) = \boldsymbol{j}^2(p,0)$ permits no existence of a one-to-one relation.

[2] This symmetry essentially stems from the equivalence between these two kinds of descriptions, and the direct implication is that we also have $f(p,0) = 1$ for the situation where $\boldsymbol{r}(x,0) = \boldsymbol{d}^2(x-x_0)$.

Thus there should exist only one term $px$ in the function $G(p,x)$, and this permits the existence of the symmetry relation between these two transformations, which will be $T^{-1}(p,x) = T^*(p,x) = e^{ib_1 px}$. We let $b_1 = 1/\hbar$, where $\hbar$ is a constant quantity with dimension $J \cdot s$. For simplicity we let $\hbar = 1$ in the following discussions unless state otherwise.

Now we get the simplest one-to-one relation, it is:

$$\boldsymbol{y}(x,0) = \int_{-\infty}^{+\infty} \boldsymbol{j}(p,0) e^{ipx} dp \quad \text{------ (7)}$$

where $\boldsymbol{y}(x,0) = e^{ip_0 x}$ and $\boldsymbol{j}(p,0) = \boldsymbol{d}(p - p_0)$. This relation mainly results from the essential symmetries involved in the discontinuous motion itself.

In order to further find how the time variable $t$ is included in the functions $\boldsymbol{y}(x,t)$ and $\boldsymbol{j}(p,t)$, we may consider the superposition of two single momentum states, namely

$$\boldsymbol{y}(x,t) = \frac{\sqrt{2}}{2} [e^{ip_1 x - ic_1(t)} + e^{ip_2 x - ic_2(t)}] \quad \text{------ (8)}$$

The corresponding position measure density is $\boldsymbol{r}(x,t) = \frac{1}{2}[1 + \cos(\Delta c(t) - \Delta p x)]$, where $\Delta c(t) = c_2(t) - c_1(t)$, $\Delta p = p_2 - p_1$. Now we let $\Delta p \to 0$, then we have $\boldsymbol{r}(x,t) \to 1$ and $\Delta c(t) \to 0$. Using the measure conservation relation we can get $dc(t) = dp \frac{p}{m} t$, then as to the nonrelativistic situation we get $c(t) = Et = \frac{p^2}{2m} t$, where

$E = \dfrac{p^2}{2m}$, is defined as the energy of the particle in the nonrelativistic domain. Thus as to any single momentum state we have the time-included formula $\boldsymbol{y}(x,t) = e^{ipx - iEt}$, and the complete one-to-one relation is:

$$\boldsymbol{y}(x,t) = \int_{-\infty}^{+\infty} \boldsymbol{j}(p,t) e^{ipx - iEt} dp \quad \text{------ (9)}$$

Since the one-to-one relation between the position description and momentum description is irrelevant to the concrete motion state, the above one-to-one relation for the free motion state with one momentum should hold true for any motion state.

In fact, there may exist more complex forms for the state functions $\boldsymbol{y}(x,t)$ and $\boldsymbol{j}(p,t)$, for example, they are not the above simple number functions but multidimensional vector functions such as $\boldsymbol{y}(x,t) = (\boldsymbol{y}_1(x,t), \boldsymbol{y}_2(x,t), \ldots, \boldsymbol{y}_N(x,t))$ and $\boldsymbol{j}(p,t) = (\boldsymbol{j}_1(p,t), \boldsymbol{j}_2(p,t), \ldots, \boldsymbol{j}_N(p,t))$. However, the above one-to-one relation still exists for every component function, and these vector functions still satisfy the above modulo square relations, namely $\boldsymbol{r}(x,t) = \sum_{i=1}^{N} |\boldsymbol{y}_i(x,t)|^2$ and $f(p,t) = \sum_{i=1}^{N} |\boldsymbol{j}_i(p,t)|^2$. These complex forms will correspond to the particles with more complex structure, say, involving more inner properties such as charge and spin etc, for example, as to the particle with spin 1/2 such as electron, we have N = 4, $\boldsymbol{r}(x,t) = \sum_{i=1}^{4} |\boldsymbol{y}_i(x,t)|^2$.

Now we can finally work out the simplest nonrelativistic evolution law of the discontinuous motion. First, as to the free motion state with one momentum, namely the single momentum state $\boldsymbol{y}(x,t) = e^{ipx - iEt}$, using the above definition of energy

$E = \dfrac{p^2}{2m}$ and including the constant quantity $\hbar$ we can easily find its nonrelativistic evolution law:

$$i\hbar \frac{\partial \psi(x,t)}{\partial t} = -\frac{\hbar^2}{2m} \frac{\partial^2 \psi(x,t)}{\partial^2 x} \quad \text{------ (10)}$$

Owing to the linearity of this equation, this evolution equation also applies to the linear superposition of the single momentum states, that is all possible free motion states. Alternatively we can say that it is the free evolution law of the discontinuous motion.

Secondly, we will consider the evolution law of the discontinuous motion under an outside potential. When the potential $U(x,t)$ is a constant $U$, the evolution equation will be

$$i\hbar \frac{\partial \psi(x,t)}{\partial t} = -\frac{\hbar^2}{2m} \frac{\partial^2 \psi(x,t)}{\partial^2 x} + U\psi(x,t) \quad \text{------ (11)}$$

Then when the potential $U(x,t)$ is related to $x$ and $t$, the above form will still hold true, namely:

$$i\hbar \frac{\partial \psi(x,t)}{\partial t} = -\frac{\hbar^2}{2m} \frac{\partial^2 \psi(x,t)}{\partial^2 x} + U(x,t)\psi(x,t) \quad \text{------ (12)}$$

For three-dimensional situation the equation will be

$$i\hbar \frac{\partial \psi(\vec{x},t)}{\partial t} = -\frac{\hbar^2}{2m} \nabla^2 \psi(\vec{x},t) + U(\vec{x},t)\psi(\vec{x},t) \quad \text{------ (13)}$$

Thus we get the simplest nonrelativistic evolution law of the discontinuous motion using the simplest one-to-one relation. We find that it is just the form of Schrödinger equation in quantum mechanics.

At last, we want to denote that the state function $\psi(x,t)$ provides a complete description of the discontinuous motion of particles. On the one hand, according to the above evolution equation, the state function $\psi(x,t)$ can be expressed by the position measure density $\rho(x,t)$ and position measure flux density $j(x,t)$, namely $\psi(x,t) = \rho^{1/2} e^{iS(x,t)/\hbar}$, where $S(x,t) = m\int_{-\infty}^{x} \frac{j(x',t)}{\rho(x',t)} dx' + C(t)$ [1]. On the other hand, the position measure density $\rho(x,t)$ and position measure flux density $j(x,t)$ can also be expressed by the state function $\psi(x,t)$, namely $\rho(x,t) = |\psi(x,t)|^2$, $j(x,t) = \frac{1}{2i}(\psi^* \frac{\partial \psi}{\partial x} - \psi \frac{\partial \psi^*}{\partial x})$. Thus there exists a one-to-one relation between $\rho(x,t)$, $j(x,t)$ and $\psi(x,t)$ when omitting the absolute phase. Since the position measure density $\rho(x,t)$ and position measure flux density $j(x,t)$ provide a complete description of the discontinuous motion of particles, the state function $\psi(x,t)$ also provides a complete description of the discontinuous motion of particles.

## The meaning of the theory of discontinuous motion

The sameness between the simplest nonrelativistic evolution equation of the discontinuous motion and the Schrödinger equation in quantum mechanics strongly

---

[1] When in three-dimensional space, the formula for $S(x,y,z,t)$ will be

$$S(x,y,z,t) = m\int_{-\infty}^{x} \frac{j(x',y,z,t)}{\rho(x',y,z,t)} dx' + C(t) = m\int_{-\infty}^{y} \frac{j(x,y',z,t)}{\rho(x,y',z,t)} dy' + C(t) = m\int_{-\infty}^{z} \frac{j(x,y,z',t)}{\rho(x,y,z',t)} dz' + C(t),$$

since in general there exists the relation $\nabla \times \{\vec{j}(x,y,z,t)/\rho(x,y,z,t)\} = 0$.

suggests what quantum mechanics describes is discontinuous motion of particles. But before reaching the definite conclusion, we need to understand the meaning of the theory of discontinuous motion. This means we must talk about measurement.

One subtle problem is what happens during a measuring process? There exist only two possibilities: one is that the measuring process still satisfies the above evolution equation of discontinuous motion or Schrödinger equation, the linear superposition of the wave function can hold all through. This possibility corresponds to the many worlds interpretation of quantum mechanics; the other is that the measuring process doesn't satisfy the above evolution equation of discontinuous motion or Schrödinger equation, the linear superposition of the wave function is destroyed due to some unknown causes. The resulting process is often called the collapse of wave function. Certainly, the above two possibilities can be tested in experiments, but unfortunately it is very difficult to distinguish them using present technology. In the following we will mainly give some theoretical considerations about them.

As to the first possibility, the discontinuous motion of particles provides the corresponding physical reality in real space-time for many worlds interpretation. The particle discontinuously moves throughout all the parallel worlds during very small time interval, or even infinitesimal time interval, and this objectively and clearly shows that these parallel complete worlds exist in the same space-time. At the same time, the measure density of the particle in different worlds, which can be strictly defined for the discontinuous motion of particle, just provides the objective origin of the measure of different worlds. Thus the visualizing physical picture for many worlds is one kind of subtle time-division existence, in which every world occupies one part of the continuous time flow, and the occupation way is discontinuous in essence, i.e. the whole time flow for each world is a dense and discontinuous instant set, and all these dense time sub-flows constitute a whole continuous time flow. In this meaning, the many worlds are the most crowded in time!

Although the above many worlds picture of particles or measuring devices can exist in a consistent way, a hard problem does appear when considering the observer, i.e. why does the observer only continuously perceive one definite world while he is still discontinuously moving throughout the many worlds? This seems to be inconsistent with one of our scientific views, according to which our perception is one kind of correct reflection of the objective world. Besides, we must solve the above observer problem in order to have a satisfying many worlds theory. This may need to resort to a theory of consciousness, but we have none up to now.

Now we turn to the second possibility. We will first find whether there exist some possible evidences for the existence of dynamical collapse in present theoretical framework. If there indeed exist some, we will then construct a preliminary theory of dynamical collapse.

## The discrete space-time and the possible origin of collapse

We have been discussing the motion of particles in continuous space-time, but it should be clearly realized that the continuity of spaced-time is just an assumption. In the nonrelativistic and relativistic domain this assumption can be applicable, and we find no essential inconsistency or paradox. But in the domain of general relativity, the motion of particle and the space-time background are no longer independent, and there exists one kind of subtle dynamical connection between them. Thus the combination of the above evolution law of discontinuous motion (or quantum mechanics) and general relativity may result in essential inconsistency, which requires that the assumption of continuous space-time must be rejected and further results in the appearance of collapse. Now let's have a close look at it.

According to general relativity, there exists one kind of dynamical connection between motion and space-time, i.e. on the one hand, space-time is determined by the motion of particles, on the other hand, the motion of particle must be defined in space-time. Then when we consider the superposition state of different positions, say position A and position B, one kind of basic logical inconsistency appears. On the one hand, according to the above evolution law of the discontinuous motion of particles (or quantum mechanics), the valid definition of this superposition requires the existence of a definite space-time structure, in which the position A and position B can be distinguished. On the other hand, according to general relativity, the space-time structure, including the distinguishability of the position A and position B, can't be pre-determined, and it must be dynamically determined by the superposition state of particle. Since the different position states in the superposition state will generate different space-time structures, the space-time structure determined by the superposition state is indefinite. Thus an essential logical inconsistency does appear!

Then what are the direct inferences of the logical inconsistency? First, its appearance indicates that the superposition of different positions of particle can't exist when considering the influence of gravity, since it can't be consistently defined in principle. It should be stressed that this conclusion only relies on the validity of general relativity in the classical domain, and is irrelevant to its validity in the quantum domain. Thus the existence of gravity described by general relativity will result in the invalidity of the linear superposition principle. This may be the origin of dynamical collapse.

Secondly, according to the definition of the superposition state of different positions of particle, its existence closely relates to the continuity of space-time, since it is required that the particle in this state should move throughout these different positions during infinitesimal time interval. Thus the nonexistence of this superposition means that infinitesimal time interval based on continuous space-time will be replaced by finite time interval, and accordingly the space-time where the particles move will display some kind of discreteness. In this kind of discrete space-time, the particle can only move throughout the different positions during finite time interval, or we can say, the particle will stay for finite time interval in any position.

Besides, it can prove that when considering both quantum mechanics and general relativity, the minimum measurable time and space size will no longer infinitesimal, but finite Planck time and Planck length. Here we will give a simple operational demonstration. Consider a measurement of the length between points A and B. At point A place a clock with mass $m$ and size $a$ to register time, at point B place a reflection mirror. When $t=0$ a photon signal is sent from A to B, at point B it is reflected by the mirror and returns to point A. The clock registers the return time. For the classical situation the measured length will be $L=\frac{1}{2}ct$, but when considering quantum mechanics and general relativity, the existence of the clock introduces two kinds of uncertainties to the measured length. The uncertainty resulting from quantum mechanics is: $dL_{QM} \geq (\frac{\hbar L}{mc})^{1/2}$, the uncertainty resulting from general relativity is: $dL_{GR} \geq \frac{Gm}{c^2}$, then the total uncertainty is: $dL = dL_{QM} + dL_{GR} \geq (L \cdot L_p^2)^{1/3}$, where $L_P = (\frac{G\hbar}{c^3})^{1/2}$, is Planck length. Thus we conclude that the minimum measurable length is Planck length $L_P$. In a similar way, we can also work out the minimum measurable time, it is just Planck time $T_P = (\frac{G\hbar}{c^5})^{1/2}$.

Lastly, we want to denote that the existence of discreteness of space-time may also imply that the many worlds theory is not right, and the collapse of wave function does exist. Since there exists a minimal time interval in discrete space-time, and each parallel world must solely occupy one minimal time interval at least, there must exist a maximal number of the parallel worlds during any finite time interval. Then when the number of possible worlds exceeds the maximal number, they will be merged in some way, i.e. the whole wave function will collapse to a smaller state space.

## A theory of dynamical collapse in discrete space-time

In this section, we will further analyze the discontinuous motion of particles in discrete space-time, and present a theory of dynamical collapse in such discrete space-time.

## A general analysis

As we know, in the discrete space-time, there exist absolute minimum sizes $T_P$ and $L_P$, namely the minimum distinguishable size of time and position of the particle is respectively $T_P$ and $L_P$. Thus in physics the existence of a particle is no longer in one position at one instant as in the continuous space-time, but limited in a space interval $L_P$ during a finite time interval $T_P$. It can be seen that this state corresponds to the instantaneous state of particle in continuous space-time, and it evidently contains no motion, but only the existence of particle. We define it as the instantaneous state of particle in discrete space-time.

Now we can get the motion state of a particle in discrete space-time from that in continuous space-time. In continuous space-time, the particle, which instantaneous state is the particle being in one position at one instant, moves throughout the whole space during infinitesimal time interval. In discrete space-time, the instantaneous state of particle turns to the particle being in a space interval $L_P$ during a finite time interval $T_P$, then the motion state of particle will naturally be that, during a finite time interval much larger than $T_P$, the particle moves throughout the whole space, which proper description is still the measure density $r(x,t)$ and measure flux density $j(x,t)$, but time-averaged. The visual physical picture of such motion will be that during a finite time interval $T_P$ the particle stays in a local region with size $L_P$, then it will still stay there or appear in another local region, which may be very far from the original region, and during a time interval much larger than $T_P$ the particle will move throughout the whole space with a certain average position measure density $\rho(x,t)$.

## The evolution of discontinuous motion in discrete space-time

In the following, we will try to find the evolution law of discontinuous motion in discrete space-time. From the above analysis, it can be anticipated that the evolution equation will be a revised Schrödinger equation, which may automatically contain the dynamical collapse process of wave function. But how should the Schrödinger equation be revised? We must find some possible clues rules.

First, since the particle does stay in a local region for a finite nonzero time interval, and appears stochastically in another local region during the next time interval, the position measure density $\rho(x,t)$ of the particle, when changed due to the invalidity of the linear superposition principle, will be essentially changed in a stochastic way, which closely relates to the concrete stay time in different stochastic region[1], and the corresponding wave function will be also stochastically changed. Thus the evolution of discontinuous motion in discrete space-time may be the combination of the deterministic linear evolution and stochastic nonlinear evolution.

Secondly, we need to further find the concrete cause resulting in the stochastic change of the position measure density $\rho(x,t)$. As we know, the evolution of wave function is determined by the Hamiltonian of the system, or the energy distribution of the system. Thus the stochastic change of the evolution may also relate to the energy distribution of the system. Now consider a simple two-level system, which state is a superposition of two static states with different energy levels $E_1$ and $E_2$, and its position measure density $r(x,t)$ will oscillate with the period of $T = \hbar/\Delta E$, where $\Delta E = E_2 - E_1$ is the energy difference. Then if the energy difference $\Delta E$ is so large that it exceeds the Planck energy $E_p$, the position measure density $r(x,t)$ will oscillate with a period shorter than the Planck time $T_P$. But as we know, the Planck time $T_P$ is the minimum distinguishable size of time in the discrete space-time, and there should be no changes during this minimal time interval. Thus the energy superposition state, in which the energy difference is larger than the Planck energy $E_p$, can't hold all through, and must gradually collapse to one of the energy eigenstates. It can be further inferred that the dynamical collapse process must happen for any energy superposition state due to the general validity of the natural law including the collapse law.

---

[1] As to the discontinuous motion in continuous space-time, the stay time of the particle in any position is zero, thus its position measure density $\rho(x,t)$ is not influenced by the stochastic motion.

Now we will work out the concrete evolution law of the discontinuous motion in discrete space-time. At first, the position of the particle will satisfy the position measure density $r(x,t)$ in the sense of time average, namely the stochastic stay position of the particle satisfies the distribution:

$$P(x,t) = |y(x,t)|^2 \quad \text{------ (14)}$$

This is the first useful rule for finding the evolution law of DSTM.

Secondly, we consider the change of position measure density $r(x,t)$ after the particle stays in a local region $L_P$ for a time interval $T$. In the first rank

approximation the change of $r(x,t)$ in this region can be written as follows after normalization:

$$r(x,t+T) = \frac{1}{A(T,T_m)}[r(x,t)+T/T_m] \quad \text{------ (15)}$$

where $A(T,T_m)$ is the normalization factor, $T_m$ is a certain time size to be determined, which may relate to the concrete motion state of the particle. It can be seen that the dynamical collapse process doesn't exist if $T_m$ is infinite. This formula will be the second useful rule.

Considering the influence of energy difference and dimensional relation, we assume $T_m = k\hbar/\Delta E$, where $\Delta E$ is the total difference of energy of the particle between the local region containing $x$ and all other regions, $k$ is a dimensionless constant. Then the above formula can be written as follows:

$$r(x,t+T) = \frac{1}{A(T,\Delta E)}[r(x,t)+T\Delta E/k\hbar] \quad \text{------ (16)}$$

Now we further consider two extreme situations:

(1). When T = 0 or $\Delta E = 0$, we have $r(x,t+T) = r(x,t)$. Then we get:

A $(0, \Delta E) = 1$, A $(T, 0) = 1$;

(2). When T → ∞ or $\Delta E \to \infty$, we have $r(x,t+T) \to 1$. Then we get:

A $(\infty, \Delta E) \to T\Delta E/k\hbar$, A $(T,\infty) \to T\Delta E/k\hbar$.

Thus we can get the formula of A(T, $\Delta E$), namely A(T, $\Delta E$) = $1+T\Delta E/k\hbar$.

Then the above formula can be written as follows:

$$r(x,t+T) = \frac{r(x,t)+T\Delta E/k\hbar}{1+T\Delta E/k\hbar} \quad \text{------ (17)}$$

when $T = T_P$ we have:

$$\mathbf{r}(x,t+T_p) = \frac{\mathbf{r}(x,t) + T_p \cdot \Delta E / k\hbar}{1 + T_p \cdot \Delta E / k\hbar} \quad \text{------ (18)}$$

or it can be written as a simpler form:

$$\Delta \mathbf{r}(x,t) = \frac{\Delta E}{kE_p + \Delta E}[1 - \mathbf{r}(x,t)] \quad \text{------ (19)}$$

where $E_p = \hbar / T_p$ is Planck energy. This formula describes the change of position measure density $\mathbf{r}(x,t)$ after a minimal time interval $T_P$ in the discrete space-time. It is the second useful rule for finding the evolution law of the discontinuous motion in discrete space-time.

Now we can give the simplest nonrelativistic evolution equation of the discontinuous motion in discrete space-time. According to the above analysis, it will be essentially one kind of revised stochastic nonlinear equation based on the Schrödinger equation. Here we assume the form of stochastic differential equation (SDE), it can be written as follows:

$$d\mathbf{y}(x,t) = \frac{1}{i\hbar} H_Q \mathbf{y}(x,t)dt + \frac{1}{2}[\frac{\mathbf{d}_{xx_N}}{\mathbf{r}(x,t)} - 1]\frac{\Delta E(x_N, \overline{x_N})}{kE_p + \Delta E(x_N, \overline{x_N})}\mathbf{y}(x,t)\frac{dt}{T_p} \quad \text{------ (20)}$$

where the first term in right side represents the linear evolution element, $H_Q$ is the corresponding Hamiltonian, the second term in right side represents the stochastic nonlinear evolution element resulting from the stochastic change of the position measure density $\mathbf{r}(x,t)$. $\mathbf{d}_{xx_N}$ is the discrete $\mathbf{d}$-function, $k$ is a dimensionless constant, $\mathbf{r}(x,t) = |\mathbf{y}(x,t)|^2$, is the position measure density, $\Delta E(x_N, \overline{x_N})$ is the total difference of energy of the particle between the local Planck cell containing $x_N$ and all other regions $\overline{x}_N$, $x_N$ is a stochastic position variable, whose distribution is $P(x_N,t) = |\mathbf{y}(x_N,t)|^2$.

Certainly, the stochastic differential equation is essentially a discrete evolution equation in physics. All of the quantities are defined relative to the Planck cells $T_P$ and $L_P$, and the equation should be also solved in a discrete way.

**Some further discussions**

Now we will give some physical analyses about the above evolution equation of the discontinuous motion in discrete space-time.

First, the linear item in the evolution equation will result in the spreading process of wave function similar to the normal evolution of wave function in quantum mechanics, while the nonlinear stochastic item in the equation will result in the localizing process of particle or collapse process of wave function. This can also be seen qualitatively. On the one hand, in the region where the position measure density is larger the accumulative stay time of the particle will be longer. On the other hand, according to the nonlinear stochastic item, the longer stay time of the particle in one region will further increase the position measure density in that region much more. Then this process is evidently one kind of positive feedback process, the particle will finally stay in a local region, and the wave function of particle will also collapse to that region when taking no account of the spreading process. Thus the evolution of the discontinuous motion in discrete space-time will be some kind of combination of the spreading process and localizing process.

Secondly, the relative strength of the spreading process and localizing process is mainly determined by the energy difference between different branches of the wave function. If the energy difference is so small, then the evolution will be mainly dominated by the spreading process. This is just what happens in the microscopic world. While if the energy difference is so large, then the evolution will be mainly dominated by the localizing process, and its display will be more like that of continuous motion. This is just what happens in the macroscopic world. The boundary of these two worlds can also be approximately estimated. The following example indicates that the energy difference in the boundary may assume $\Delta E \approx 7 Gev$, and the corresponding collapse time will be in the level of $10^{-6} s$.

Thirdly, if the particle finally stays in a local region during the evolution, the localizing probability of the particle, or the collapse probability of the wave function in a local region is just the initial position measure density of the particle in that region, i.e. the collapse probability satisfies the Born rule in quantum mechanics. In fact, the stochastic evolution of the discontinuous motion in discrete space-time satisfies the Martingale condition[7]. This can be seen from the following fact, i.e. during every step the position measure density $\boldsymbol{r}$ satisfies the equation:

$$P(\boldsymbol{r}) = \boldsymbol{r} P(\boldsymbol{r}+\boldsymbol{a}) + (1-\boldsymbol{r}) P(\boldsymbol{r}-\boldsymbol{b}) \quad \text{------ (21)}$$

where $P(\boldsymbol{r})$ is the probability of $\boldsymbol{r}$ turning into one in one local region, namely the probability of the particle localizing in a local region, $\boldsymbol{a} = \dfrac{\Delta E}{kE_p + \Delta E}(1-\boldsymbol{r})$, $\boldsymbol{b} = \dfrac{\Delta E}{kE_p + \Delta E}\boldsymbol{r}$. Moreover, the solution of this equation is $P(\boldsymbol{r}) = \boldsymbol{r}$. This means that the localizing probability of the particle in one region is just the initial position measure density of the particle in that region.

Lastly, the existence of the discontinuous motion in discrete space-time may help to tackle the well-known time problem in quantum gravity[8], and a complete theory of quantum gravity may be formulated based on it. Since as to the discontinuous motion in discrete space-time, the local position state of a particle will be the only proper state and real physical existence. During a finite time interval $T_P$ the particle can only be limited in a local space interval $L_P$, thus there does not exist any essential superposition of different positions at all. The superposition of the wave function can only be found in the meaning of time average, thus the essential inconsistency of the superposition of different space-time in some theories of quantum gravity, which results from the existence of the essential superposition of the wave function, will naturally disappear. The physical picture based on the discontinuous motion in discrete space-time will be that at any instant (during a finite time interval $T_P$) the structure of space-time determined by the existence of the particle (in a local space interval $L_P$) is definite or "classical", while during a finite time interval much larger than $T_P$ but still small enough, it will be stochastically disturbed by the stochastic appearance of the particle. This kind of stochastic fluctuation may be the real quantum nature of space-time and matter.

## An example

In the following, as an example we will analyze the evolution of a simple two-level system, and quantitatively show that the evolution of the discontinuous motion in discrete space-time will indeed result in the dynamical collapse process of wave function.

Suppose the initial wave function of the particle is $\psi(x,0) = a(0)^{1/2} \psi_1(x) + b(0)^{1/2} \psi_2(x)$, which is a superposition of two static states with different energy levels $E_1$ and $E_2$. These two static states are located in separate regions $R_1$ and $R_2$ with the same size.

Since the energy of the particle inside the region of each static state is the same, we can consider the spreading space of both static states as a whole local region, i.e. we can directly consider the difference of the energy $\Delta E = E_2 - E_1$ between these two states. Besides, we only consider the nonlinear stochastic item in the evolution equation of the discontinuous motion in discrete space-time, since the linear item only results in a phase factor, and doesn't influence our conclusion. Through some mathematical calculations we can work out the density matrix of the two-level system:

$$\rho_{11}(t) = a(0) \quad \text{----- (22)}$$

$$\rho_{12}(t) \cong [1 - \frac{(\Delta E)^2}{2k^2 \hbar E_p} t] \sqrt{a(0) b(0)} \quad \text{------ (23)}$$

$$\rho_{21}(t) \cong [1 - \frac{(\Delta E)^2}{2k^2 \hbar E_p} t] \sqrt{a(0) b(0)} \quad \text{------ (24)}$$

$$\rho_{22}(t) = b(0) \quad \text{------ (25)}$$

It is evident that these results confirm the above qualitative analysis definitely, i.e. the evolution of the discontinuous motion in discrete space-time indeed results in the collapse of the wave function, and the distribution of the collapse results satisfies the Born rule in quantum mechanics. Besides, we also get the concrete collapse time for two-level system: $t_c \approx 2k^2 \frac{\hbar E_p}{(\Delta E)^2}$ [1].

**The appearance of continuous motion in the macroscopic world**

The above analysis has indicated that, when the energy difference between different branches of the wave function is large enough, say, for the macroscopic situation[2], the linear spreading of the wave function will be greatly suppressed, and the evolution of the wave function will be dominated by the localizing process. Thus a macroscopic object will be always in a local position, and it can only be still or continuously move in space in appearance. This is just the display of continuous motion in the macroscopic world.

Furthermore, we can show that the evolution law of continuous motion can also be derived from the evolution of the discontinuous motion in discrete space-time. In fact, some people have strictly given the demonstration based on the similar revised quantum dynamics[9][10]. Here we only give a simple explanation using the Ehrenfest theorem, namely $\frac{d<x>}{dt} = \frac{1}{m}<p>$ and $\frac{d<p>}{dt} = <-\frac{\partial U}{\partial x}>$. As we have demonstrated, for a macroscopic object its wave function will no longer spread, thus the average items in the theorem will represent the effective description quantities for the continuous motion of the macroscopic object. Then the evolution law of continuous motion can be naturally derived in such a way, the result is: $\frac{dx}{dt} = \frac{p}{m}$, the definition of the momentum, and $\frac{dp}{dt} = -\frac{\partial U}{\partial x}$, the motion equation.

---

[1] The similar result has also been obtained by Percival[11], Hughston[12] and Fivel[13] from different points of views, and discussed by Adler et al[14].

[2] The largeness of the energy difference for a macroscopic object results mainly from the environmental influences such as thermal energy fluctuations.

## Conclusions

In this paper, we present a new interpretation of quantum mechanics, according to which what quantum mechanics describes is discontinuous motion of particles. We formulate a theory of discontinuous motion of particles in continuous space-time, and demonstrate that its simplest nonrelativistic evolution equation is just the Schrödinger equation in quantum mechanics. Whereas space-time will be essentially discrete when considering gravity, we further present a theory of discontinuous motion of particles in discrete space-time. We show that the evolution of such motion will naturally result in the dynamical collapse process of the wave function, and this collapse will finally bring about the appearance of continuous motion in the macroscopic world. This gives a unified realistic picture of the microscopic and macroscopic world.